\documentclass[12pt, a4paper]{article}
\usepackage[utf8]{inputenc}

\usepackage{times}           
\usepackage[scaled=0.92]{helvet}  

\usepackage[top=2.5cm, bottom=3.5cm, 
            left=2.5cm, right=2.5cm,
            headheight=17pt]{geometry}
\newcommand{\hangpara}[1]{%
    \par\hangindent=1.2em\hangafter=1\noindent#1\par}
    
\usepackage{parskip}                   
\usepackage{xcolor}
\usepackage{makecell}
\usepackage{lineno}      

\usepackage{amsmath}  
\usepackage{textcomp} 
\usepackage[none]{hyphenat}
\usepackage{lineno}

\usepackage{graphicx}
\usepackage{booktabs} 
\usepackage[font=small, labelfont=bf]{caption}

\usepackage[super,numbers,sort&compress]{natbib}
\usepackage[hypertexnames=false,           
            pdfencoding=auto]{hyperref}    
\makeatletter
\renewcommand\@biblabel[1]{\textsuperscript{#1}}  
\patchcmd{\thebibliography}
  {\section*{\refname}}
  {\section*{\refname}\setlength{\itemsep}{0pt} \setlength{\labelwidth}{0em} \setlength{\leftmargin}{2em}}{}{}
\makeatother

\makeatletter
\renewcommand{\@maketitle}{%
  \newpage
  \null
  \vspace*{-0pt} 
  \begin{center}
    \vspace{24pt}
    {\fontsize{18}{22}\bfseries\selectfont \@title \par} 
    \vspace{24pt} 
    {\fontsize{15}{16}\selectfont \@author \par} 
    \vspace{24pt} 
    {\fontsize{15}{16}\selectfont \@date}
  \end{center}
  \vspace{20pt} 
}
\makeatother

\begin{document}
\setlength{\parindent}{2em} 
\setlength{\parskip}{0\baselineskip} 

\title{\textbf{Turbulence Physics Governs a Scaling Law for the Machine-Learning Predictability Ceiling in Chaotic Flow}}

\author{Jiashun Guan,\textsuperscript{1,2,3,*} Haoyang Hu,\textsuperscript{2} Yunxiao Ren,\textsuperscript{4,$\dagger$} 
Michael S. Triantafyllou,\textsuperscript{5} and Dixia Fan\textsuperscript{3,$\ddagger$}}
\date{\today} 
\maketitle

\hangpara{\noindent \textsuperscript{1} ~ School of Aeronautics, Universidad Politécnica de Madrid, Madrid 28040, Spain}

\hangpara{\noindent \textsuperscript{2} ~ School of Mechanics and Engineering Science, Peking University, Beijing 100871, China}

\hangpara{\noindent \textsuperscript{3} ~ School of Engineering, Westlake University, Hangzhou 310024, China}

\hangpara{\noindent \textsuperscript{4} ~ College of Engineering, Peking University, Beijing 100871, China}

\hangpara{\noindent \textsuperscript{5} ~ Department of Mechanical Engineering, Massachusetts Institute of Technology, Cambridge, MA 02139, USA}
\vspace{3ex}
\hangpara{\noindent \textsuperscript{~} ~ To whom correspondence should be addressed: \\ 
Email:\\ 
$^*$  jiashun.guan@upm.es; \\
$^\dagger$  renyx@pku.edu.cn; \\
$^\ddagger$  fandixia@westlake.edu.cn}
\vspace{16ex}

\clearpage

\setlength{\parindent}{0pt}
\setlength{\parskip}{1\baselineskip} %
\textbf{For centuries, the intrinsic chaos of unsteady fluid motion has stood as a formidable barrier to long-term forecasting. While machine learning (ML) has recently emerged as a transformative paradigm for predicting flow evolution, it encounters a pervasive yet unexplained ``performance wall": an inevitable deterioration in accuracy as the forecast horizon extends. Here, we demonstrate that this deterioration is not a deficiency of model architecture, no matter how state-of-the-art, but a fundamental constraint imposed by the underlying system, which can be understood through turbulence theory established decades ago. In the setting of bluff body flow, a canonical phenomenon for spatiotemporal complexity in fluid mechanics, we reveal a scaling law governing the deterioration of ML predictability, derived from a Kolmogorov-inspired framework and validated through high-fidelity simulations. Our findings establish a closed loop between the predictability ceiling and its interpretation, bridging the gap between transparent physical theories and modern black-box inference. More broadly, this work provides a theoretical compass for constructing trustworthy ML in complex dynamical systems across the physical sciences.}

{\fontsize{16}{16}\selectfont \textbf{Introduction}}\\
In the development of human societies, predicting unsteady flow has long been a fundamental challenge, sometimes even influencing critical decisions that shape civilizations. The ability to predict fluid flows also has broad implications for natural and engineered systems, spanning scales from the microscopic flow within our arteries\cite{scarselli2023} to the large-scale organization in the universe\cite{alexakis_large-scale_2024,bi_accurate_2023}. Heraclitus said that ``No man ever steps in the same river twice,'' reflecting the unsteadiness of flow, whose multiscale, multifractal, and even multiphase characteristics introduce difficulties in prediction. To address these challenges, theories established since the early nineteenth century \cite{navier1823,stokes1845}, together with observations that include experiments \cite{poiseuille1846,reynolds1883} and simulations \cite{rogallo1984,chen1998}, have been developed as the two primary approaches for understanding and predicting the chaotic behavior of fluid flow.

With the advent of the machine learning era \cite{lecun_deep_2015, brunton2020,vinuesa_transformative_2023,karniadakis2021physics}, notable advances have been achieved using these nonlinear and high-dimensional tools \cite{raissi2020hidden, bae_scientific_2022, fan_reinforcement_2020}. Machine learning offers a pathway for linking present and future states by capturing the nonlinear dynamical evolution in phase space, whose inherently fractal nature lies beyond our intuitive understanding \cite{pathak2018,guan_vortex-induced_2025, Kim2025,barthel_harnessing_2023}. Despite the remarkable progress driven by machine learning in chaotic flows, a rigorous return to first principles is required to establish the theoretical boundaries of these methods. Deepening our understanding of where ML meets classical physics is a key path toward deriving the insights necessary for the next generation of robust, physics-consistent models.

A sobering precedent can be found in numerical weather prediction: despite a century of paradigm shifts and exponential growth in computing power, the practical forecast horizon remains limited to roughly $1$--$2$ weeks \cite{bauer_quiet_2015}. This barrier underscores that predictability is an inherent dynamical behavior of the nonlinear system, rather than a mere reflection of algorithmic sophistication. This fundamental limit was revealed by Lorenz's pioneering work \cite{lorenz1963}, which demonstrated, in an extremely simple form of math model, that the deterministic system can exhibit sensitivity to infinitesimal uncertainties (widely known as the butterfly effect). Lorenz’s legacy shifted the scientific focus from the pursuit of computational brute force to the recognition of a hard physical ceiling that no algorithm can transcend.

In this spirit, we adopt the flow past a circular cylinder as an ideal laboratory for probing system predictability. This configuration, though geometrically simple, harbors rich physical complexity. As a canonical phenomenon in fluid mechanics for symmetry breaking and spatiotemporal chaos, its dynamics are universal and ubiquitous, governing phenomena ranging from planetary-scale island wakes to the aerodynamics of skyscrapers and launch vehicles \cite{williamson_vortex_1996}.

Here, we predict the temporal evolution of the drag coefficient, $C_d$, in cylinder flow by utilizing surface pressure signals as inputs to a deep learning architecture. By identifying the most informative spatial locations on the cylinder surface, our models reveal an exponential scaling law governing the deterioration of prediction performance as the forecast horizon extends. We demonstrate that this scaling is theoretically rooted in the linear growth of errors within the underlying fluid system, driven by an inverse-cascade process that transfers errors from small to large scales. To understand this behavior, we develop a top-down theory inspired by Kolmogorov scaling analysis and validate it through extensive direct numerical simulations (DNS) involving $\mathcal{O}(10^6)$ central processing unit (CPU) hours and over 8 TB of data. Our results establish a closed loop between observation and theory, defining the predictability ceilings of chaotic fluid systems and providing a physics-tethered foundation for the reliability of black-box machine learning in complex dynamics.

\begin{figure}[t]
    \linespread{1.} 
    \centering
    \includegraphics[width=1\linewidth]{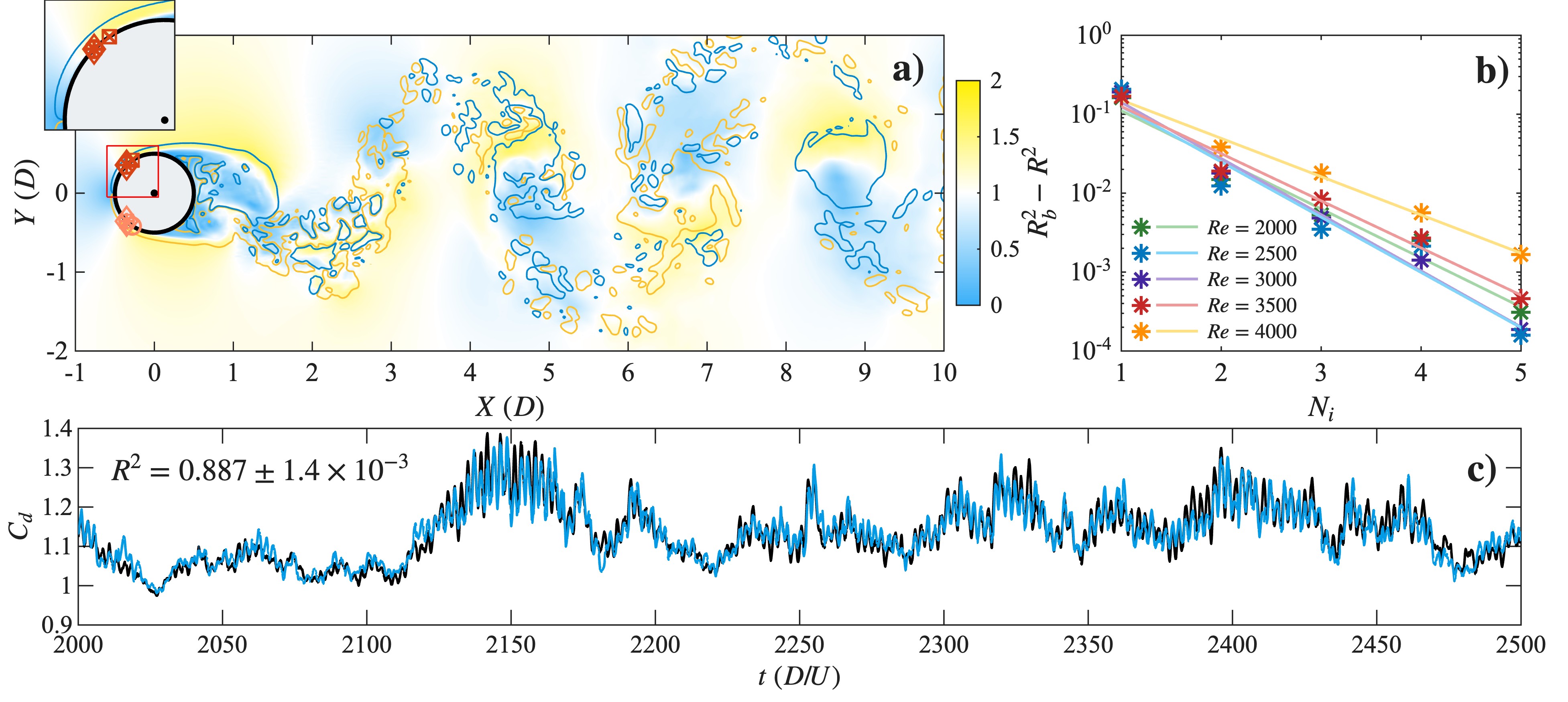}
    \caption{\textbf{Drag prediction for the cylinder flow.} \textbf{a)} An instantaneous visualization of cylinder wake at $Re=4000$. The yellow and blue isopleths represent the vortex criterion $Q=\pm 1 U^2/D^2$, respectively. The streamwise velocity is shown by the contour plot, where the color ranges from blue for low speeds, through white (corresponding to the free stream speed, $U=1$), to yellow for high speeds. The first two best-performing locations are marked with red and orange symbols. Circle, diamond, asterisk, cross, and square symbols denote the results at Reynolds numbers from $2000$ to $4000$, respectively (details in Methods). The inset is the zoomed-in view of the first best-performing locations. The first two best-performing locations align with the key physical characteristic of the cylinder wake: they are symmetrically positioned  $45^\circ$ (relative to the $X$-axis) upstream of the separation points ($\approx 90^\circ$) \cite{jiang_separation_2020, guan_vortex-induced_2025}, corresponding to the most informative locations in cylinder wake dynamics. \textbf{b)} Performance approaches the ceiling value $R_b^2$ exponentially with increasing $N_i$. $R^2_b$ is estimated by the fit $R^2=R^2_b-C_1\exp(C_2N_i)$, where $C_1$ and $C_2$ are parameters of linear regression on the semi-logarithmic axes. Symbols represent the mean values averaged over five predictions. \textbf{c)} Prediction ($t_p=1 D/U$) versus DNS time series of $C_d$ in the testing set at $Re=4000$ (prediction: blue line; DNS: black line). The $R^2$ metric reached $0.887\pm 1.4\times 10^{-3}$ (mean $\pm$ standard deviation over five independent predictions). \\}
    \label{fig:f1}
\end{figure}
{\fontsize{16}{16}\selectfont \textbf{Results}}\\
\textbf{Drag Prediction for the Cylinder Flow}\\
We consider the prototypical chaotic fluid system (Fig. \ref{fig:f1}a): the flow passes a circular cylinder at a moderate Reynolds number $Re=UD/\nu$, where $U$, $D$, and $\nu$ are the free stream velocity, cylinder diameter, and kinematic viscosity, respectively. At moderate Reynolds numbers (specifically $2000\leq Re \leq 4000$, as considerated in this paper), the cylinder flow exhibits complex dynamics: boundary layer separation gives rise to shear-layer instabilities that amplify disturbances and generate a sequence of chaotic vortices \cite{wei_secondary_1986,williamson_vortex_1996}. The three-dimensional velocity vector $\mathbf{u}=(u,v,w)$ and pressure $p$ of the flow are governed by the incompressible Navier–Stokes equations (NS equations):
\begin{subequations}
\begin{equation}
     \frac{\partial \mathbf{u}}{\partial t} + (\mathbf{u} \cdot \nabla) \mathbf{u}  = -\nabla p + \frac{1}{Re} \nabla^2 \mathbf{u},
\end{equation}
\begin{equation}
    \nabla \cdot \mathbf{u} = 0,
\end{equation}
\label{eq:navier_stokes}
\end{subequations}
and the DNSs are implemented by the open source spectral element code Nek5000\cite{nek5000_webpage_2008}. The computational domain ($30D\times20D\times3D$ in the $X$, $Y$, and $Z$ directions) is discretized into 22,740 spectral elements with $7^{th}$-order accuracy (for numerical details, see Methods). A circular array of 32 equally spaced sensors was positioned on the cylinder surface within the $Z=0$ plane, recording the pressure signal at a sampling rate of $100$ per time unit ($U/D$). Pressure signals are processed using the discrete wavelet transform (DWT)\cite{sundararajan2015}, with the $7^{th}$-level coefficients (corresponding to the frequency band $f\in(100/2^8, 100/2^7]$, see Methods) selected for the following training. For the time-series prediction of drag coefficient $C_d$, we employed a deep learning approach, in which a fully connected neural network with five hidden layers constructs the complicated mapping from the processed pressure signals to the drag value at a forecast horizon $t_p$, serving as the input and output, respectively. The prediction performance is quantified by the $R^2$ metric:
\begin{equation}
    R^2=1-\frac{1}{N} \sum^{N}_{i=1}\frac{(\hat C_d -C_d )^2}{\operatorname{Var}(C_d)},
    \label{eq:r2}
\end{equation}
where $\hat C_d$ and $ C_{d}$ represent the predicted and true (DNS) drag coefficient, respectively; $N$ denotes the size of the testing set used for training; and $\operatorname{Var}(\cdot)$ denotes the variance operator. In physical terms, $R^2$ reflects how much better the model performs than simply using the mean values as predictions.  Details about the prediction model are provided in the Methods. 

The optimized combination of pressure signals is probed by a greedy algorithm. Fixing $t_p=1 (D/U)$, we initialize the input with a single signal and then add the best-performing signal (ranked by $R^2$) at each optimization iteration. As the number of signals in the input $N_i$ increases, the prediction performance exponentially approaches its ceiling $R^2_b$ (a fitted value; see Fig.\ref{fig:f1}b). This exponential convergence to $R^2_b$ indicates that the global dynamic of the cylinder wake can be captured from only a few locations\cite{guan_vortex-induced_2025}; specifically, using merely three optimized pressure signals as input approaches about $90\%$ of the $R^2_b$ performance (Fig.\ref{fig:f1}b). These locations of optimized signals are also consistent with the most informative locations identified by mutual information\cite{shannon1949mathematical} (see Methods). As shown in Fig. \ref{fig:f1}c for an example drag prediction at $Re=4000$, $N_i$ is set to 3. This value is maintained for all subsequent predictions to keep consistency in this study.

\begin{figure}[t]
    \linespread{1.} 
    \centering
    \includegraphics[width=0.66\linewidth]{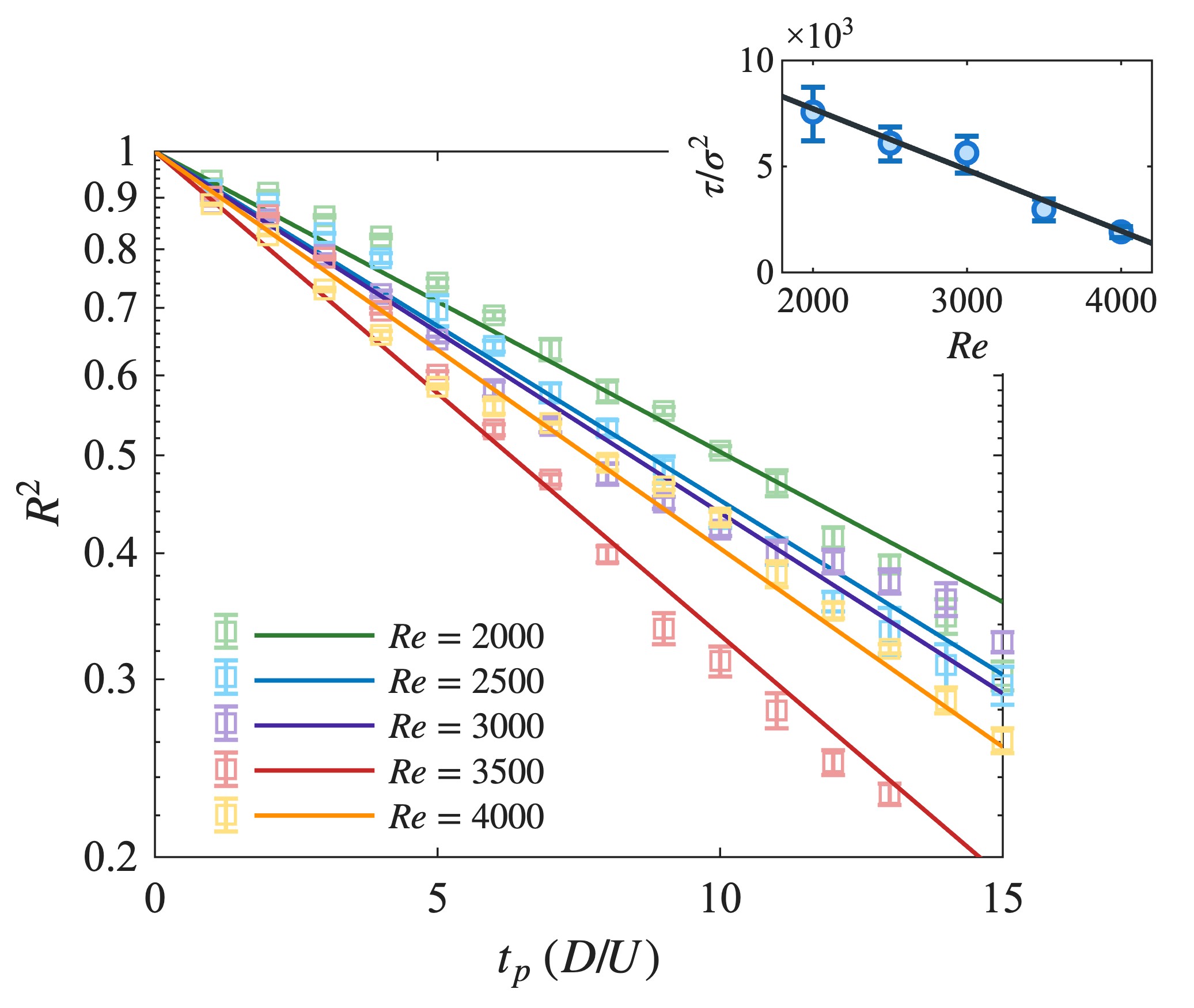}
    \caption{\textbf{The exponential scaling law.} Error bars represent the mean $\pm$ standard deviation from five independent predictions. Solid lines are fitted by Eq. \ref{eq2:exp}. The inset shows the dependence between the standard deviation $\sigma$, the timescale $\tau$, and Reynolds number $Re$. Error bars on $\tau/\sigma^2$ denote the $99\%$ confidence interval, where the uncertainty in $\sigma$ is estimated by subsampling with a subsample size of $N_{data}\Delta t_s$, and the uncertainty in $\tau$ is estimated from the exponential fit. Here, $N_{data}$ and $\Delta t_s$ represent the number of data points and data sampling interval (see Methods), respectively. The gray line in the insert is a linear fit to $\tau/\sigma^2$, with a regression $R^2_{fit}=0.96$.\\}
    \label{fig:f2}
\end{figure}
\textbf{Exponential Scaling Law for Predictability}\\
Originating from the nonlinear term in the NS equations (Eq. \ref{eq:navier_stokes}), the dynamics of cylinder flow at moderate Reynolds numbers is chaotic, and can be conceptually associated with an intricate set of the dynamical system, i.e., a chaotic attractor \cite{lanford_strange_1982}. Consequently, the pronounced sensitivity of such systems to perturbations fundamentally limits the accuracy ceiling of time-series prediction, which results in progressively deteriorating prediction performance as $t_p$ increases. We observed this deterioration both qualitatively and quantitatively: the prediction performance with optimized inputs follows an exponential scaling law across a range of Reynolds numbers (Fig. \ref{fig:f2}),
\begin{equation}
    R^2=\exp(-t_p/\tau),
    \label{eq2:exp}
\end{equation}
where the slope of the scaling laws determines the timescale $\tau$, characterizing how long the predictability is retained. Larger $\tau$ corresponds to a slower loss of predictability. The exponential scalings at different Reynolds numbers in Fig.\ref{fig:f2} do not collapse onto a single curve, indicating a Reynolds number dependence of the timescale $\tau$. Moreover, the timescale $\tau$ is directly influenced by $\sigma=\sqrt{\operatorname{Var}(C_d)}$, which acts as a normalization factor in the definition of $R^2$ (Eq.\ref{eq:r2}). To account for this intrinsic dependence, the normalized timescale $\tau/\sigma^2$ is considered, enabling a consistent comparison across different fluctuation levels of the system. The results show that this normalized timescale decreases with increasing Reynolds number, as illustrated in the inset of  Fig.\ref{fig:f2}. Even with variations in $\tau/\sigma^2$, the exponential scaling law exhibits robustness and captures the characteristic behavior of prediction deterioration in a chaotic fluid system. Furthermore, this exponential scaling has an asymptotic limit at large $t_p$, indicating that long-term predictability of chaos becomes infeasible, reminiscent of Poincaré’s findings in the three-body problem \cite{poincare1890troiscorps}.

\textbf{Does the Observed Scaling Originate from Lyapunov Growth?}\\
The exponential form of the observed scaling at first glance seems to echo Lyapunov growth. From a dynamical systems perspective, a chaotic attractor exhibits exponential divergence of nearby trajectories under infinitesimal perturbations, usually termed Lyapunov growth. This mechanism arises from the linearized dynamics around a trajectory. In contrast, the observed exponential scaling law describes the deterioration of prediction accuracy, which arises in part from finite errors introduced by the neural network; their finite amplitudes involve nonlinear mechanisms that cannot be neglected. Therefore, despite the apparent similarity in exponential behavior, the observed scaling does not originate from Lyapunov growth.

The exponential scaling law of machine learning performance can be interpreted in terms of two complementary asymptotic limits corresponding to small and large values of $t_p$. Following the definition of $R^2$ (see Eq. \ref{eq:r2}), we write $R^2=1-\delta$, where $\delta$ denotes a scaled error term.  In practice, $ R^2$ may become negative; however, negative values only indicate performance worse than the known strategy, the mean-value predictor ($R^2=0$), which can simply improve any negative-$R^2$ results. Thus, in the large-$t_p$ limit, the asymptotic behavior of the exponential scaling is consistent with $R^2 \rightarrow 0$, corresponding to the state in which the fluid fields become totally uncorrelated, and the energy of their difference reaches a saturated upper bound ($\delta=1$) \cite{boffetta_predictability_2001,ge_production_2023}. In the opposite limit $\delta \sim 0$, a Taylor expansion of $\ln(1-\delta)$ gives:
\begin{equation}
\ln(R^2) = -\delta - \frac{\delta^2}{2} - \frac{\delta^3}{3} - \cdots.
\label{eq4}
\end{equation}
Neglecting the higher-order terms, we obtain the linear approximation $ \ln(R^2) \simeq -\delta$. Therefore, if $\delta$ grows approximately linearly with $t_p$ at early times, $\ln(R^2)$ becomes proportional to $t_p$, explaining the observed exponential scaling.

\begin{figure}[t]
\linespread{1.}
    \centering
    \includegraphics[width=1\linewidth]{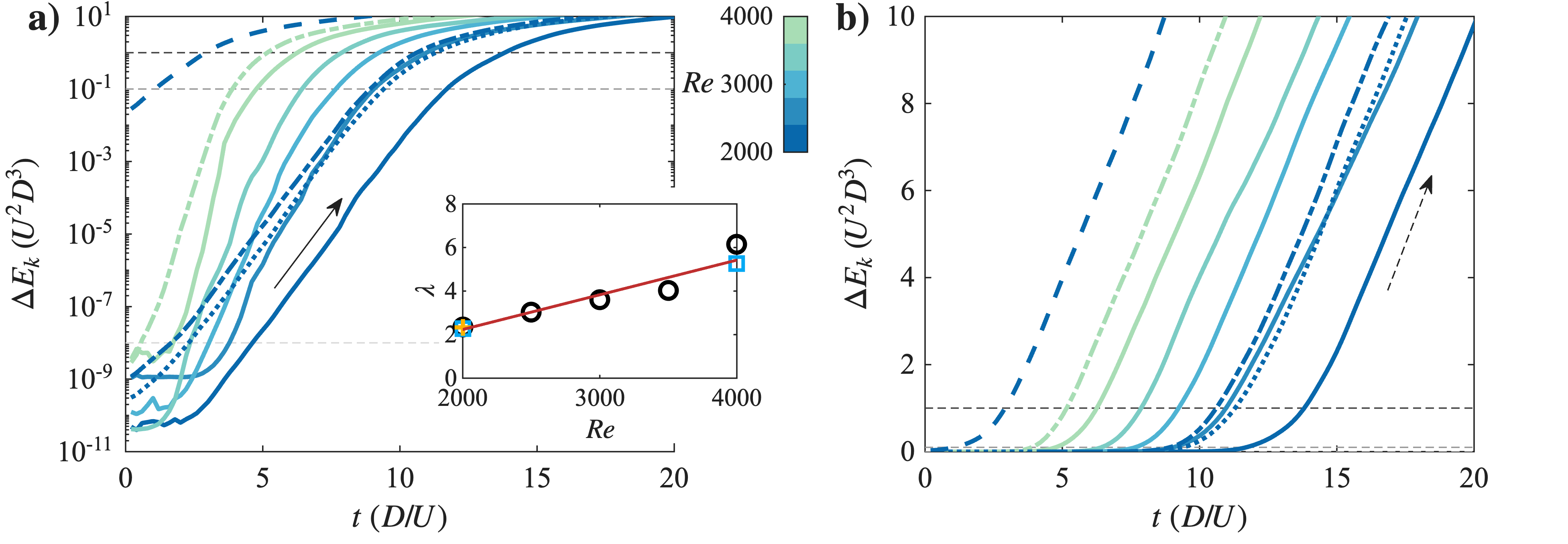}
    \caption{\textbf{The exponential growth (linear mechanism) and linear growth (nonlinear mechanism) of error.} \textbf{a)} Temporal evolution of perturbation energy $\Delta E_k$ ($U^2D^3$) in semi-logarithmic coordinates. The inset shows the Lyapunov exponent $\lambda$, obtained from the slope of $\ln(\Delta E_k)$ in the exponential growth regime, with the red line indicating a linear regression. Data in the panel \textbf{a} and the inset correspond to different initial perturbations: solid lines (panel \textbf{a}) and black symbols (inset) for $\mathbf{u}'=(0,0,10^{-7}\mathcal{N})$, where $\mathcal{N}$ is the standard Gaussian operator; dash-dotted lines and blue squares for $\mathbf{u}'=(0,10^{-5}v_0,0)$; the dotted line and yellow plus sign for $\mathbf{u}'=(0,0,10^{-5}w_0)$. The dashed line represents a finite-amplitude perturbation with energy $||\mathbf{u}'||^2> 10^{-2}$. Horizontal gray lines demarcate four growth regimes: initial modulation ($\Delta E_k<10^{-8}$), exponential growth ($10^{-8}<\Delta E_k<10^{-1}$), transition ($10^{-1}<\Delta E_k<10^{0}$, and linear growth ($10^{0}<\Delta E_k$). \textbf{b)} Temporal evolution of perturbation energy $\Delta E_k$ in linear-linear coordinates. Arrows in this figure mark the two growth regimes: the solid arrow denotes exponential growth, while the dashed arrow indicates linear growth.\\}
    \label{fig:f3}
\end{figure}

To mimic the error term $\delta$ introduced by neural networks and trace its influence, we intervene in the statistically stationary state of the fluid system $\mathbf{u}_0=(u_0,v_0,w_0)$ by an infinitesimal perturbation $\mathbf{u}'$. In the DNS implementation, the perturbation field $\mathbf{u}'$ is impossible to be infinitesimal \cite{liao_noise-expansion_2025}. Therefore, we maintain it at a low amplitude, with energy $||\mathbf{u}'||^2\ll 10^{-8}$. To eliminate the stochastic influence of the perturbation’s specific flow structure, $\mathbf{u}'$ is constructed in two ways:  the Gaussian noise and the scaled velocity field, i.e., $\mathbf{u}'=(0,10^{-5}v_0,0)$, or $(0,0,10^{-5}w_0)$. Considering the spatially anisotropic behavior of the cylinder wake, 10 independent DNS realizations are implemented for ensemble averaging to reduce statistical variability. We track the temporal evolution of the velocity field $\mathbf{u}=\mathbf{u}_0+\mathbf{u}'$ and also the ensemble averaged perturbation energy $\Delta E_k(t)=\langle||\mathbf{u}-\mathbf{u}_0||^2\rangle$, as shown in Fig. \ref{fig:f3}.  

The perturbation initially undergoes exponential growth (Fig. \ref{fig:f3}a), a process explained by linear stability theory \cite{Schmid2001Stability} and quantitatively characterized by the Lyapunov exponent $\lambda$ (the inset of Fig. \ref{fig:f3}a) \cite{nastac_lyapunov_2017}. These low-amplitude perturbations have insufficient energy to instantaneously break the correlation between $\mathbf{u}(t)$ and $\mathbf{u}_0(t)$ within large-scale flow structures. Nevertheless, once amplified to a finite magnitude through exponential growth, the perturbation energy saturates at small scales and proceeds with `inverse-cascade' \cite{boffetta_predictability_2001,boffetta_chaos_2017} to large scales, triggering the transition to linear growth (Fig. \ref{fig:f3}b). The growth rates of linear growth are observed to be constant across the Reynolds numbers in this study. Notably, a finite initial perturbation, mimicking the prediction error introduced by a neural network, causes the system to bypass exponential growth and transition directly to linear growth with the same growth rate (dashed lines in Fig. \ref{fig:f3}). This indicates that the linear error accumulation is insensitive to how the error is introduced, and instead represents a robust feature at finite error amplitudes. This linear growth leads to the observed exponential scaling law and establishes a predictability ceiling for the underlying fluid system.

To reinforce the interpretation of the observed exponential scaling law, we introduce a counterexample here. As the Reynolds number decreases to $Re<1000$, the wake transitions toward quasi-periodic shedding, i.e., the \textit{disordered three dimensionalities regime}\cite{williamson_vortex_1996}, where the abrupt onset of linear error growth gives way to a more gradual progression. Consequently, the prediction performance will no longer follow the exponential scaling law (see an example at $Re= 500$ in Methods); in turn, this supports the interpretation that the exponential scaling of ML prediction performance stems from the linear error growth.

\begin{figure}[t]
\linespread{1.}
    \centering
    \includegraphics[width=1\linewidth]{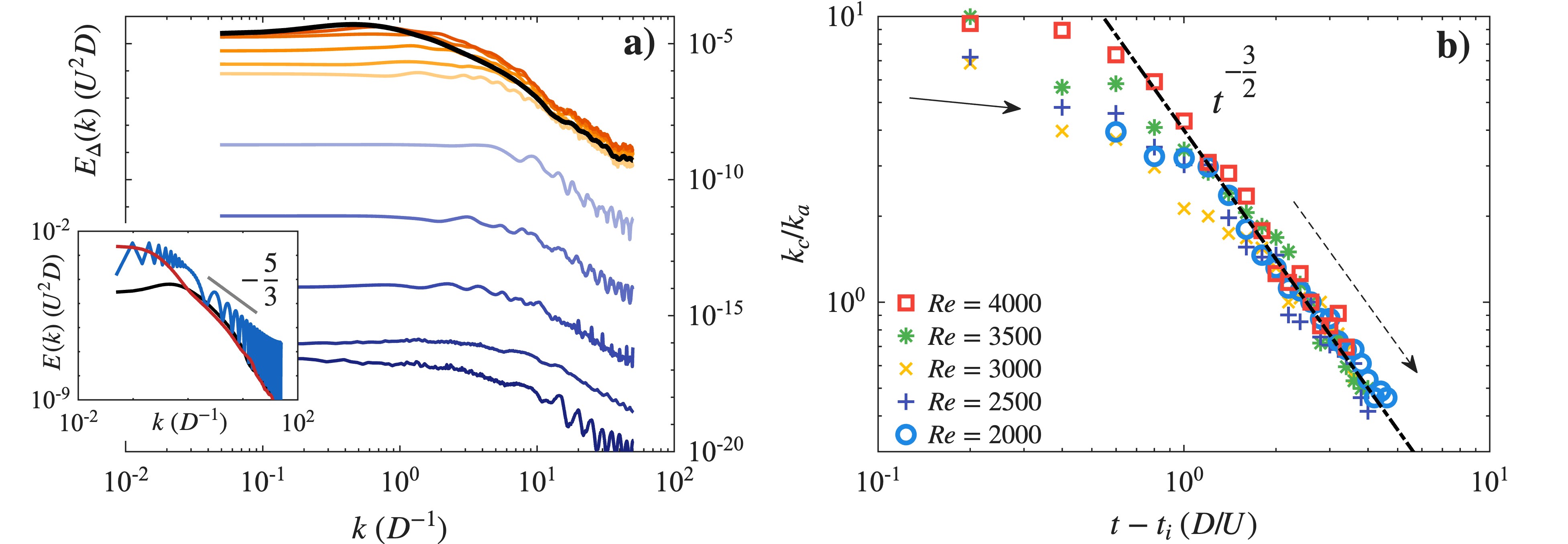}
    \caption{\textbf{The inverse-cascade of error.} \textbf{a)} Temporal evolution of the error spectra at $Re=4000$: energy spectrum of the $Y$-dependent perturbation velocity $w'=w-w_0$ on the $X=2D$ plane ($X=1D$, $3D$ also examined for robust validation), averaged along the $Z$-direction and 10 independent DNS realizations. In exponential growth (purple lines), the spectra begin at $t=0.4D/U$ with a time interval of $0.8D/U$; in linear growth (orange lines), they begin at $t=4.8D/U$ with the same interval. This temporal evolution corresponds to that of $\Delta E_k (t)$ in Fig. \ref{fig:f3}a. The black line represents $E_\Delta(k,t)/E(k)=\eta=0.8$ for numerically estimating $k_s$ ($\eta=0.9$ has also been checked for robustness). The Kolmogorov energy spectra for $u$, $v$, and $w$ on the $X=2D$ plane are shown in the inset as blue, red, and black lines, respectively, ensemble-averaged along the $Z$-direction over 750 velocity fields from 10 independent DNS realizations. \textbf{b)} Temporal evolution of critical wavenumber $k_c$ at moderate Reynolds numbers. The function $k_c(t)$ are scaled by $k_a=k_c(t_i+2.8)$, where $t_i$ is the initial time when $k_c$ becomes detectable once reaching $\eta E(k)$. The black dash-dotted line denotes the theoretical scaling result $k_c(t) = \alpha \varepsilon^{-1/2}t^{-3/2}$. }
    \label{fig:f4}
\end{figure}
\textbf{A Top-down Interpretation of the Linear Growth}\\
The predictability ceiling imposed by linear growth of error is rooted in the dynamics of turbulent flow described by the NS equations. To uncover this phenomenon from a first-principles perspective, we trace back to the intervened DNS experiments but analyze them through the evolution of perturbation energy in spectral–temporal space. Initially, in the exponential growth regime, the error spectra exhibit self-similar growth, exemplified by the purple lines at $Re=4000$ in Fig. \ref{fig:f4}a. However, the Kolmogorov energy spectrum bounds this self-similar growth \cite{kolmogorov1941local}: the error spectrum first reaches the Kolmogorov $-5/3$ scaling in the high wavenumber (small scales) and propagates to the low wavenumber (large scales), as illustrated by orange lines in Fig. \ref{fig:f4}a. The wavenumber $k_c(t)$ is identified as the value at which $E_\Delta(k,t)/E(k)=\eta \sim 1$, denoting a critical value below which the error spectrum is saturated and velocity fields $\mathbf{u}$ and $\mathbf{u}_0$ are uncorrelated. This decorrelation process at an inertial-range scale $\mathcal{L}=1/k$ occurs over a characteristic time proportional to the eddy turnover time $\tau_e$ at that scale \cite{frisch1995turbulence}: $\tau_e\sim\varepsilon^{-1/3}\mathcal{L}^{2/3}$, where $\varepsilon$ is the turbulent energy dissipation rate. From this scaling analysis, the critical wavenumber is derived to evolve as $k_c(t) = \alpha \varepsilon^{-1/2}t^{-3/2}$, with $\alpha$ as a proportional coefficient. The DNS results in Fig. \ref{fig:f4}b validate this theoretical derivation. 

The perturbation energy, dominantly constituted by energy from wavenumbers exceeding the critical value $k_c$, is calculated by integrating the Kolmogorov energy spectrum $E(k)=C\varepsilon^{2/3}k^{-5/3}$:
\begin{equation}
    \Delta E_k=\int_{k_c}^{+\infty} C\varepsilon^{2/3}k^{-5/3}dk=A\varepsilon t,
    \label{eq:linear}
\end{equation}
where $A$ is a dimensionless constant describing the linear growth rate and $C$ is the Kolmogorov constant. This linear growth is sustained while the critical wavenumber $k_c$ lies within the inertial range, where the integrand in Eq.\ref{eq:linear} remains valid. Once the inverse cascade reaches large scales, the Kolmogorov $-5/3$ scaling no longer holds for the integrand, at which stage the perturbed field becomes decorrelated from the original field \cite{boffetta_predictability_2001, ge_production_2023}.

{\fontsize{16}{16}\selectfont \textbf{Discussion}}\\
The numerically validated derivation of linear growth provides a first-principles understanding of the exponential scaling law observed in machine learning prediction performance. This exponential deterioration indicates a predictability ceiling that remains invariant regardless of algorithmic sophistication. Instead, it is a fingerprint of the underlying high-dimensional chaotic system, whose dynamics actively amplify errors, perturbations, and uncertainties\cite{vela-martin_uncertainty_2025}. Consequently, although the governing dynamics remain deterministic, future phase-space trajectories cannot be reliably resolved by finite-dimensional prediction approaches and are better described through probability density distributions\cite{jimenez_perronfrobenius_2023,guan2025onsetmetastableturbulencepipe}.

By rigorously closing this observation–theory loop, we move beyond the traditional criticism of ML as ``black boxes" and the debate between those seeking interpretable architectures and those favoring explainable outcomes\cite{rudin2019stop}. We propose that ML performance is tightly ``physics-tethered" to the inherent dynamical properties of the underlying system. This tethering perspective provides a diagnostic lens through which to build theory-constrained trust and gain deeper insight into the conceptual foundations of next-generation physics-consistent ML that is grounded in the governing principles of nature\cite{karniadakis2021physics, liang2022advances, li2023trustworthy}. 

Crucially, this synergy is not a one-way constraint but opens a reciprocal gateway for exploring systems whose emergent behaviors remain inscrutable. This reciprocity can be framed in the context of complex dynamical systems -- from turbulent flows to plasma and astrophysical dynamics -- where ML can extend the reach of empirical, numerical, and theoretical exploration by revealing high-dimensional relationships that are difficult to access directly, while established physical theories, in turn, endow learning systems with structure, constraints, and interpretive depth\cite{brunton2020}. By pinpointing exactly where the model’s performance ceiling lies, our findings do more than delineate the epistemic boundaries of ML; they serve as a compass for guiding human intellectual effort toward the underlying physical bedrock, where our scientific intuition can work in concert with increasingly artificial-intelligence-driven discovery \cite{raissi2020hidden, davies2021advancing,klowden2026mathematical, lu2026towards}.

{\fontsize{16}{16}\selectfont \textbf{Acknowledgements}}\\
Computational resources are supported by superserver@js. The support from the National Natural Science Foundation of China is acknowledged (Grant No. 12572280). We thank Prof. Javier Jiménez, Prof. Bernd R. Noack, and Prof. Themistoklis Sapsis for insightful and fruitful discussions.

\clearpage
\setlength{\parindent}{0pt}
{\fontsize{20}{16}\selectfont \textbf{Methods}}
\renewcommand{\thetable}{S\arabic{table}} 
\renewcommand{\thefigure}{S\arabic{figure}}  
\renewcommand{\theequation}{S\arabic{equation}}  
\setcounter{figure}{0}
\setcounter{equation}{0}

\begin{figure}[h]
\linespread{1.}
    \centering
    \includegraphics[width=0.8\linewidth]{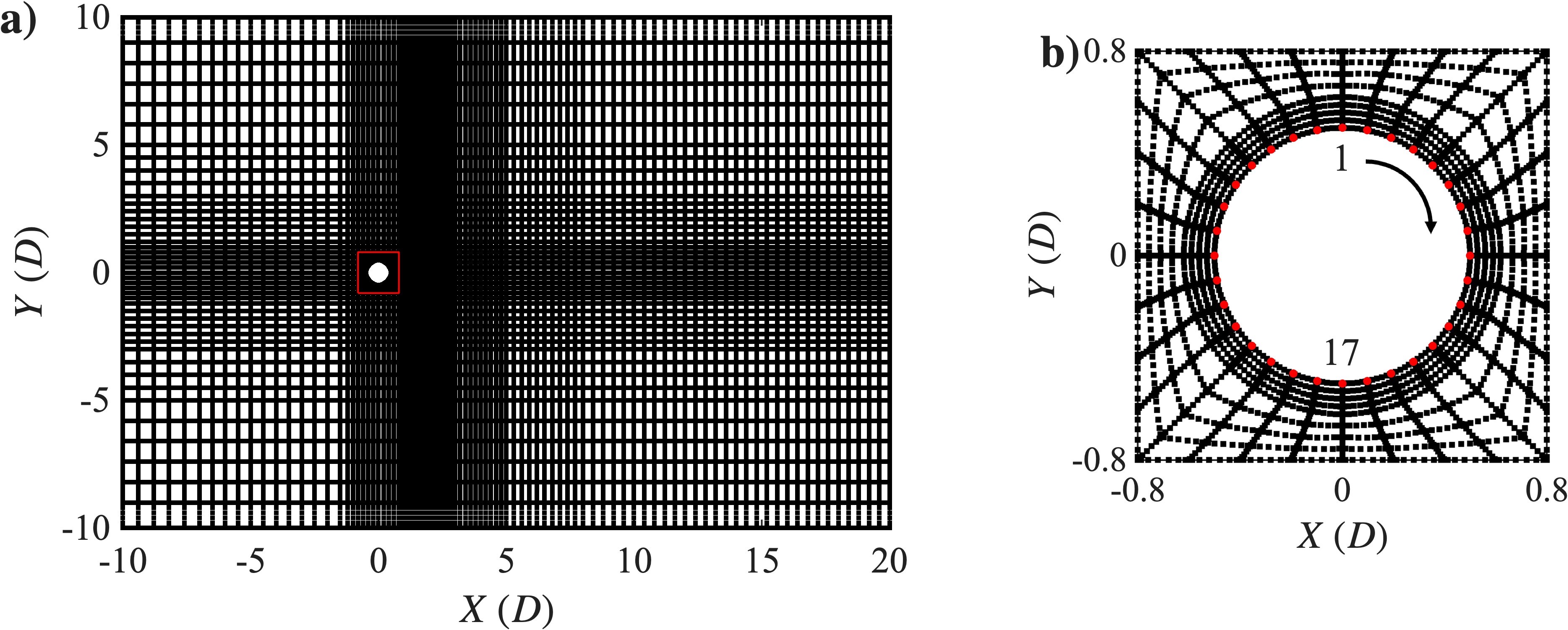}
    \caption{\textbf{Computational mesh.} \textbf{a)} Full-domain view for $4548$ spectral elements in the $X$–$Y$ plane. The region outlined in red is magnified in panel \textbf{b)}. \textbf{b)} Detailed view of the mesh near the cylinder. The red points represent the locations of pressure sensors. The index $pt$ of pressure signals begins at the top position (where $pt=1$) and increases in the clockwise direction. }
    \label{fig:mesh}
\end{figure}

{\fontsize{16}{16}\selectfont \textbf{Setup of Direct Numerical Simulations }}\\
Direct numerical simulations (DNS) of the incompressible Navier-Stokes equations (Eq.\ref{eq:navier_stokes}) are performed by the open source spectral element code Nek5000\cite{nek5000_webpage_2008}. The computational domain is established in Cartesian coordinates, taking a space of $30D\times20D\times3D$ in the $X$, $Y$, and $Z$ directions, respectively. The mesh (Fig. \ref{fig:mesh}) is locally refined near the cylinder and discretized into 22,740 spectral elements. A fixed inlet velocity (dirichlet-type) and an open (outflow) boundary conditions\cite{nek5000_webpage_2008} are applied in the $X$ direction to ensure a time-invariant Reynolds number.  
\begin{figure}[!h]
    \linespread{1.} 
    \centering
    \includegraphics[width=0.55\linewidth]{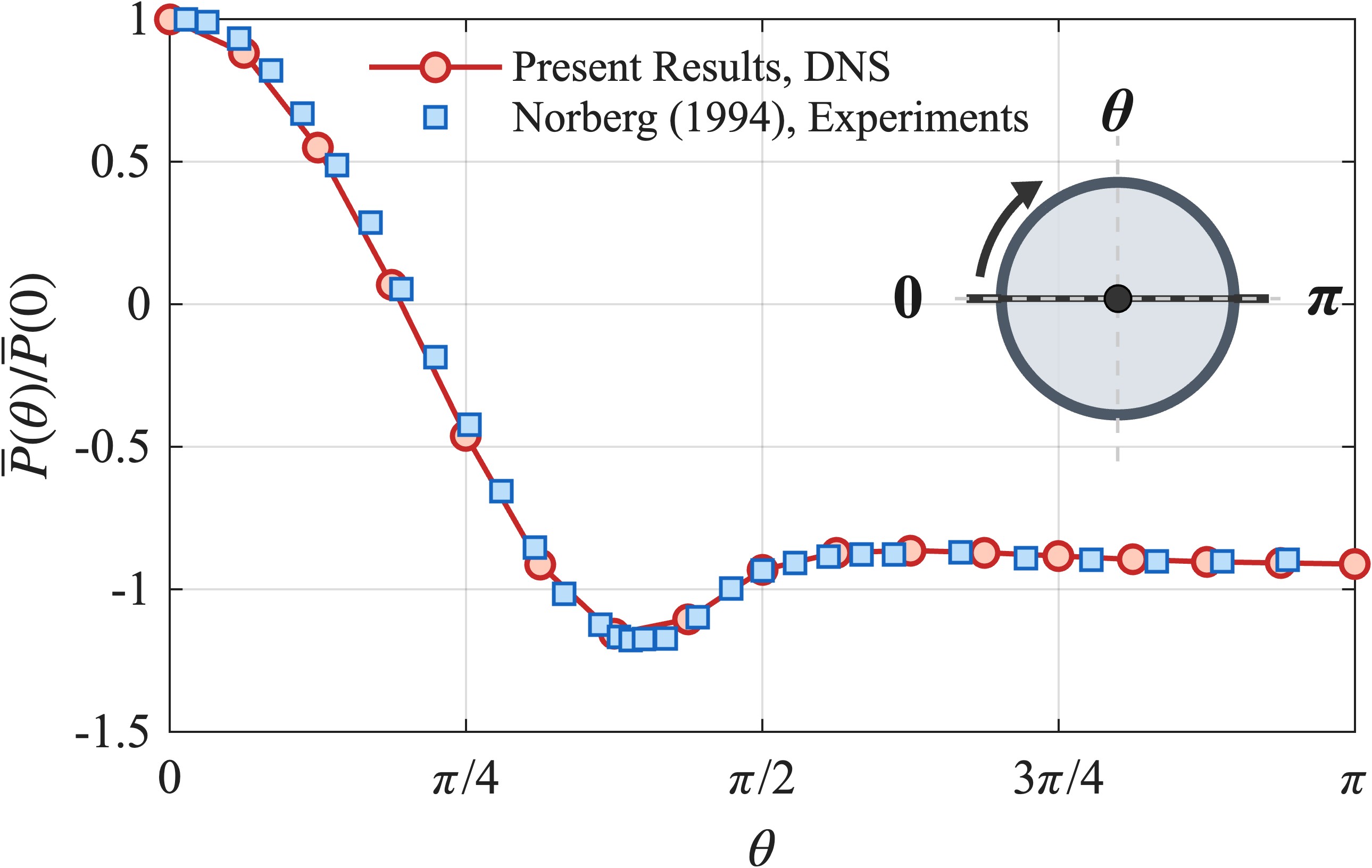}
    \caption{\textbf{Validation of numerical results.} The present numerical results are compared with the experimental data \cite{Norberg1994} at $Re=4000$, the highest Reynolds number considered in this study. $\overline{P}(\theta)$ denotes the pressure averaged over the total simulation time $T_{max}$.}
    \label{fig:vali}
\end{figure}
Periodic conditions are applied in the $Y$ and $Z$ directions. The time stepping is set to $ 0.0005 D/U$ for $Re>500$ and $0.001D/U$ for $Re=500$, resulting in a CFL (Courant-Friedrichs-Lewy) number of approximately $0.5$ to ensure numerical stability. The numerical results are validated against the experimental data\cite{Norberg1994} at the highest Reynolds number considered in this study, $Re=4000$, as shown in Fig. \ref{fig:vali}. Pressure signals at locations shown in Fig.\ref{fig:mesh}b are sampled every $\Delta t_s=0.01 D/U$. The total simulation interval (Tab.\ref{tab:dns}) for each Reynolds number covers hundreds of vortex shedding cycles to generate adequate data for the following training. The numerical intervention experiments are conducted as summarized in Tab.\ref{tab:dnsstability}, providing the data used for the theoretical analysis in Fig.\ref{fig:f3} and Fig.\ref{fig:f4}.

\begin{table}[h]
\linespread{1.}
\centering
\begin{tabular}{cccc}
\toprule
$Re$ & $T_{max}$ ($D/U$) &Training Set, $T_{train}$ ($D/U$) & Testing Set ($D/U$) \\
\midrule
4000 & 2500 & 0$\sim$2000 & $T_{max}-500 \sim T_{max}$ \\
3500 & 3000 & 0$\sim$2500 & $T_{max}-500 \sim T_{max}$ \\
3000 & 3500 & 0$\sim$3000 & $T_{max}-500 \sim T_{max}$ \\
2500 & 3500 & 0$\sim$3000 & $T_{max}-500 \sim T_{max}$ \\
2000 & 4000 & 0$\sim$3500 & $T_{max}-500 \sim T_{max}$ \\
500  & 6000 & 0$\sim$5500 & $T_{max}-500 \sim T_{max}$ \\
\bottomrule
\end{tabular}
\caption{\textbf{Total simulation intervals $T_{max}$ and train-test data splitting.}}
\label{tab:dns}
\end{table}

\begin{table}[h]
\linespread{1.}
\centering
\begin{tabular}{ccccc}
\toprule
$Re$ & The formation of $\mathbf{u}' $  & $N_{case}$ & $T_{max}$ ($D/U$) & Saved Fields \\
\midrule
$4000$ & $(0,0,10^{-7}\mathcal{N})$; $(0,10^{-5}v_0,0)$  & $20$ & $15$; $15$. & $75$; $75$.  \\
$3500$ & $(0,0,10^{-7}\mathcal{N})$ & $10$ & $15$ & $75$   \\
$3000$ & $(0,0,10^{-7}\mathcal{N})$ & $10$ & $16$ & $80$   \\
$2500$ & $(0,0,10^{-7}\mathcal{N})$ & $10$ &    $19$ & $95$ \\
$2000$ & \makecell{$(0,0,10^{-7}\mathcal{N})$; $(0,10^{-5}v_0,0)$;\\$(0,0,10^{-5}w_0)$; $(0,0.05v_0,0)$.} & $40$ & $22$; $20$; $20$; $10$. &  $110$; $100$; $100$; $50$. \\
$500$  & $(0,0,10^{-7}\mathcal{N})$ & $10$ & $60$ &  $300$  \\
\bottomrule
\end{tabular}
\caption{\textbf{The initial conditions and statistical parameters of numerical intervention experiments.} $T_{max}$ here denotes the simulation intervals for each of the $N_{case}$ cases.}
\label{tab:dnsstability}
\end{table}

{\fontsize{16}{16}\selectfont \textbf{Details of Drag Prediction }}\\
\textbf{Setup of the Machine Learning Model}
\begin{figure}[t]
    \linespread{1.} 
    \centering
    \includegraphics[width=0.6\linewidth]{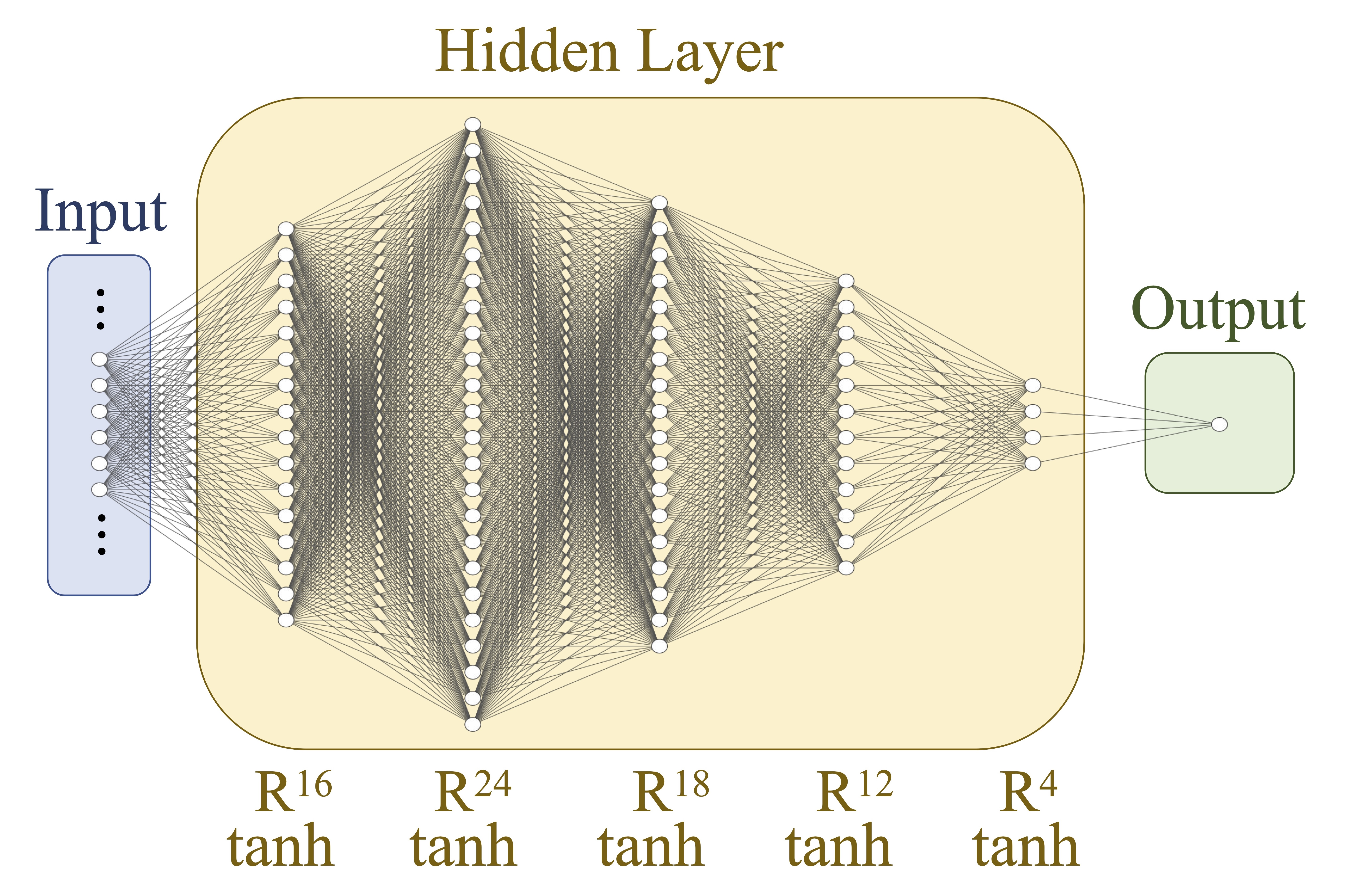}
    \caption{\textbf{Architecture diagram of the fully connected neural network (FCNN).} $R^{i}$ represents the neuron count of each fully connected layer.}
    \label{fig:fcnn}
\end{figure}
\begin{figure}[!h]
    \linespread{1.} 
    \centering
    \includegraphics[width=0.9\linewidth]{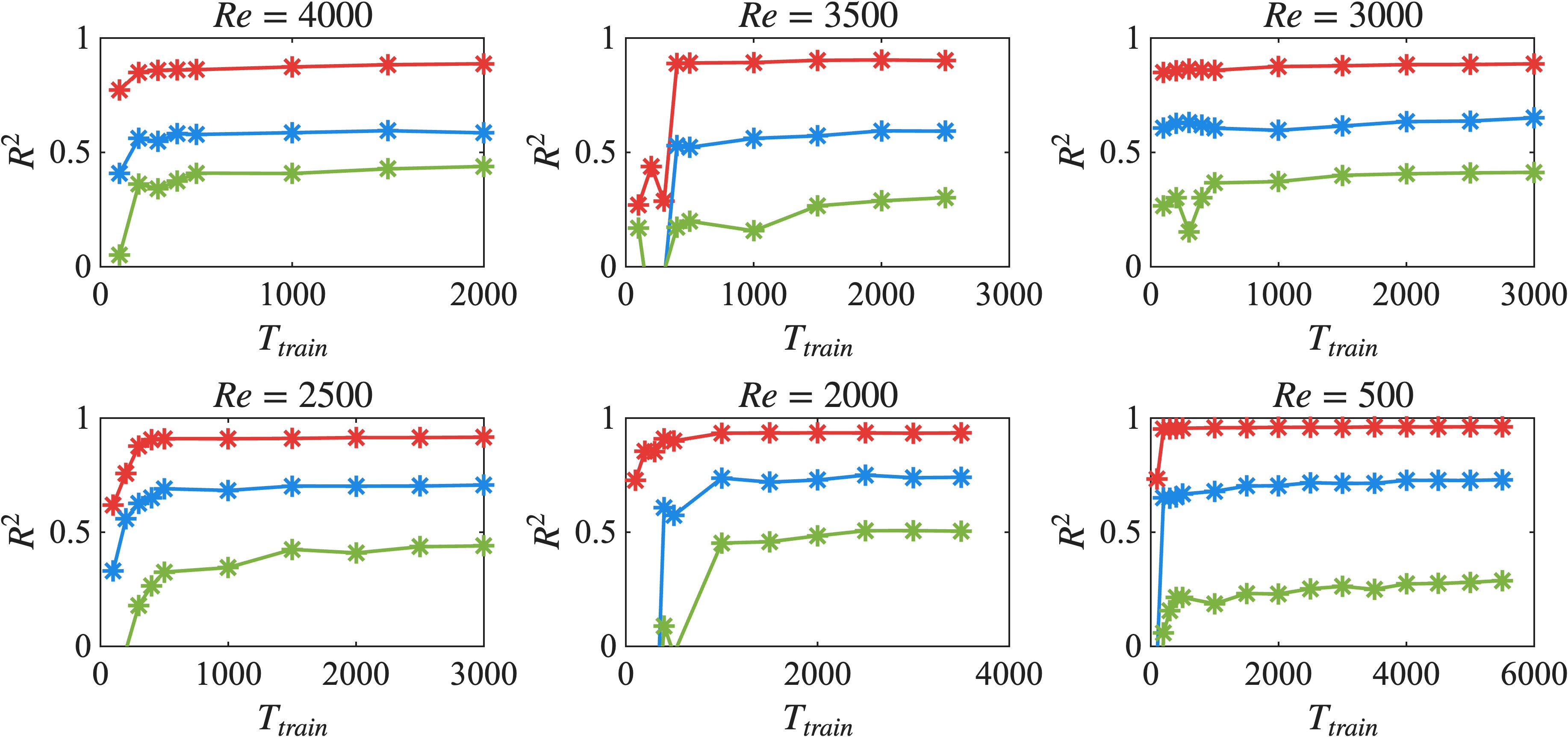}
    \caption{\textbf{Verification of Training Set Adequacy.} $T_{train}$ represents the temporal length of the training set, and the testing set remains consistent during different $T_{train}$. Three forecast horizons $t_p=1$, $5$, and $10~D/U$ are shown by red, blue, and green symbols.}
    \label{fig:ts}
\end{figure}

A fully connected neural network (FCNN) is employed for $C_d$ prediction. The FCNN architecture is structured with hidden layers of size 16-24-18-12-4 between the input and output layers (Fig.\ref{fig:fcnn}). The input is linearly normalized between $-1$ and $1$ by a min-max scaler before the training to stabilize the training process. After each hidden layer, a hyperbolic tangent function $f(x)=\tanh(x)$ is applied to provide nonlinearity to the network. The output of this network is the prediction of the drag coefficient after $t_p$ time units: $C_d(t+t_p)$. For each Reynolds number, the dataset is divided into training and testing sets: the last 50,000 samples (corresponding to the final $500$ time units shown in Tab. \ref{tab:dns}) are used as the testing set. 
\begin{figure}[t]
    \linespread{1.} 
    \centering
    \includegraphics[width=0.95\linewidth]{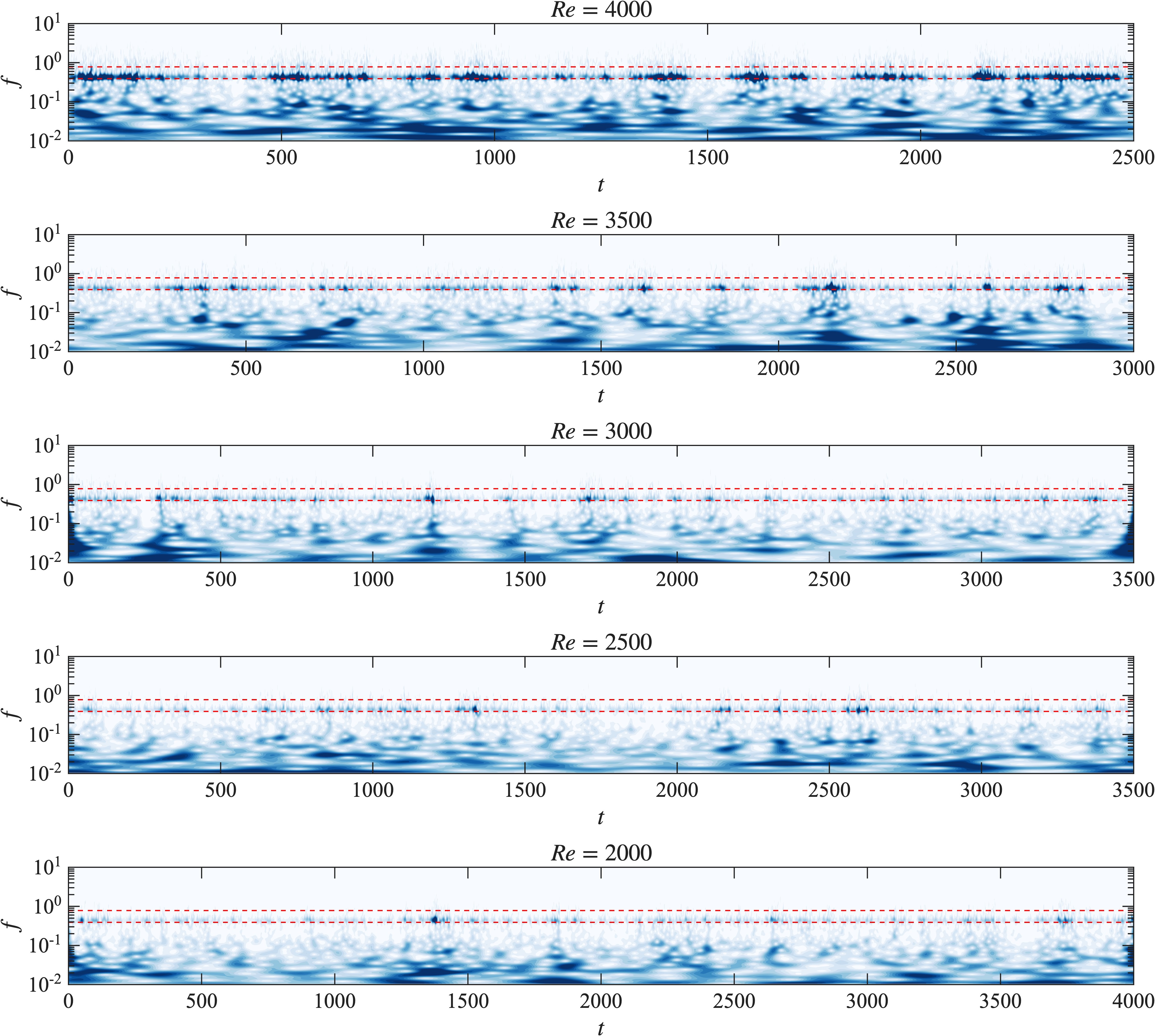}
    \caption{\textbf{Continuous wavelet transform processed $C_d$ signal.} The contours represent amplitudes of the frequencies $f$, linearly spanning the range $[0, 0.025]$. Regions outlined by red dashed lines represent the frequency band corresponding to the $7^{th}$-level coefficients in the discrete wavelet transform used for training.}
    \label{fig:wave}
\end{figure}
The AdamW optimizer\cite{loshchilov2019adamw} is employed with an initial learning rate of $10^{-3}$ and a weight decay parameter of $5\times10^{-4}$. A cosine annealing learning rate scheduler with warmup is adopted to prevent overfitting. The total number of training epochs is 30, of which the first 3 are used for warmup. We adopt the mean absolute error (MAE) as the loss function:
\begin{equation}
    MAE=\frac{1}{N}\sum_{i=1}^{N}|\hat{C}_d(t)-C_d(t)|,
\end{equation}
where $\hat C_d$ denotes the predicted drag coefficient. The model performance is evaluated via the $R^2$ metric (Eq.\ref{eq:r2}), and the results in this study correspond to the average $R^2$ over five independent training runs with identical model settings. The adequacy of the training set is verified at all Reynolds numbers for three forecast horizons: $t_p=1$, $5$, and $10$ $D/U$ (Fig.\ref{fig:ts}).

\textbf{Data Processing for Input}\\
To analyze the time-frequency dynamics of the drag coefficient $C_d$, the continuous wavelet transform (CWT) using the Morse wavelet\cite{olhede2012generalized} is performed on the time series of $C_d$ obtained via DNS, as shown in Fig.\ref{fig:wave}. These time-frequency contours (Fig.\ref{fig:wave}) indicate that peaks at high frequency $f=0.4\sim0.5$ are dominant over the entire time period for all Reynolds numbers. Based on the CWT analysis, a more computationally efficient method, the discrete wavelet transform (DWT) \cite{sundararajan2015}, is applied to decompose the pressure signals to the $7^{th}$-level corresponding to the dominant frequency range. The time window $T_w$ required by the $7^{th}$ level DWT decomposing is given as $T_{w}= 2^n\Delta t_s=1.28 D/U$, where $n$ denotes the $n^{th}$-level DWT. The processed $7^{th}$-level DWT coefficients $\gamma$ of every pressure signal and their time derivative $d\gamma/dt\approx(\gamma(t)-\gamma(t-\Delta t_s))/\Delta t_s$ are combined to construct the input column vector of size $2\times N_i$, where $N_i$ is the number of pressure signals used (e.g. Fig.\ref{fig:f1}b).

\begin{figure}[h]
    \linespread{1.2} 
    \centering
    \includegraphics[width=1\linewidth]{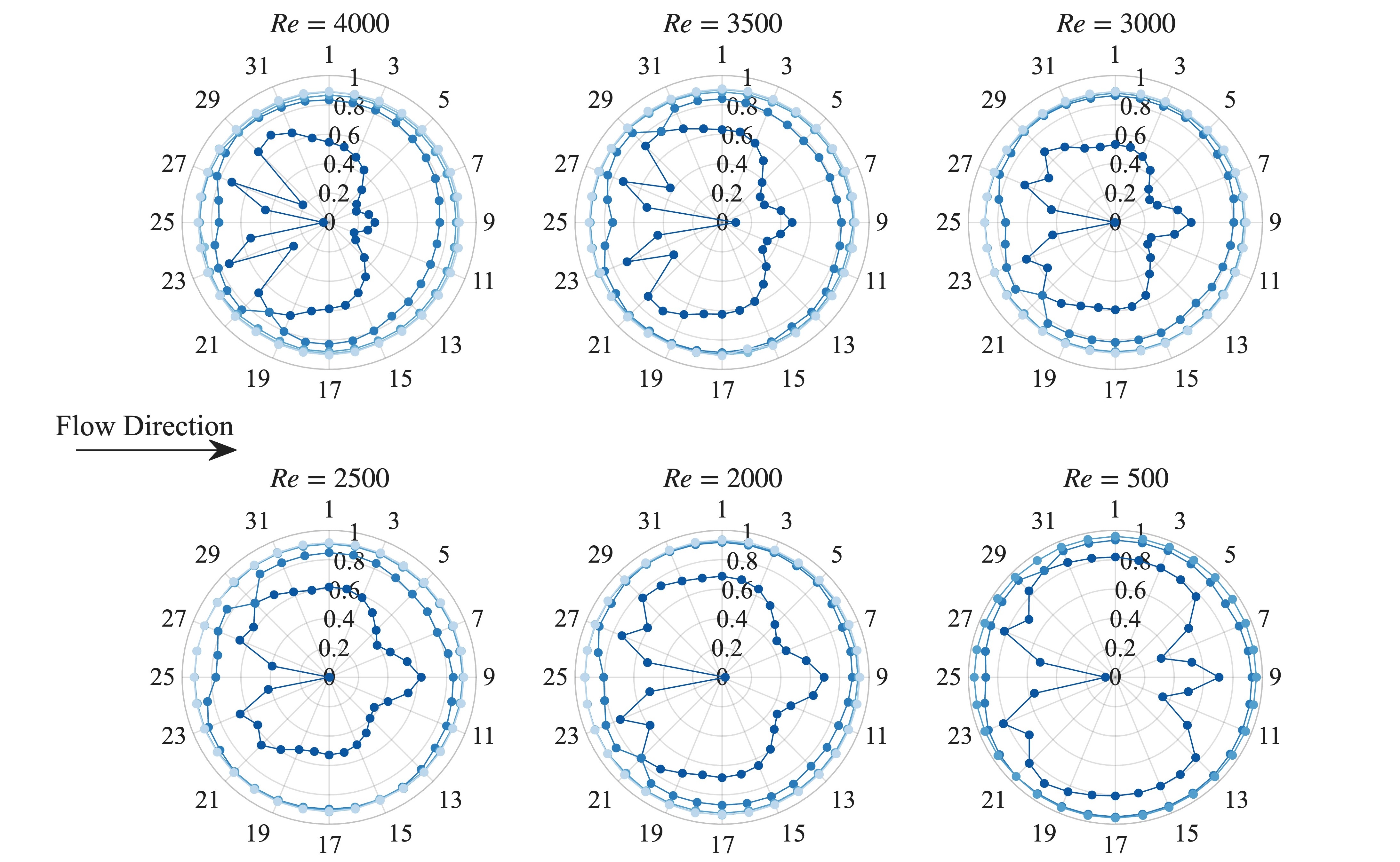}
    \caption{\textbf{Optimization iterations across 32 pressure signals. } The darkest blue symbols represent the first iteration. The azimuthal direction and radial direction denote the pressure signal index $pt$ and $R^2$, respectively.}
    \label{fig:opt_pt}
\end{figure}

\begin{table}[h]
    \linespread{1.} 
    \centering
    \begin{tabular}{ccc}
    \toprule
       $Re$  & $pt$ with $T_{train}$ & $pt$ with $T_{train}-500$ \\
    \midrule
       $4000$ &  $\quad 20,29,22,28,21 \quad$ & $\quad 23,27,29 \quad$ \\
       $3500$ &  $\quad 30,21,28,22,29 \quad$ & $\quad   30,21,23 \quad$ \\
       $3000$ &  $\quad 21,29,23,27,30 \quad$  & $\quad 21,30,23 \quad$ \\
       $2500$ &  $\quad 29,21,26,31,24 \quad$   & $\quad 29,21,27 \quad$ \\
       $2000$ &  $\quad 21,30,24,3,18 \quad$  & $\quad 21,30,24 \quad$ \\
       $500$  &  $30,21,28$  & $30,21,28$ \\
       \bottomrule
    \end{tabular}
    \caption{\textbf{Optimization iteration results.} The optimized signal locations are also validated by a shorter training set with $T_{train}-500$.}
    \label{tab:opt_pt}
\end{table}

\textbf{Optimization Iteration for the Input}\\
The number of possible combinations of the input column vector reaches $2^{32} > 10^9$ types, making it practically infeasible to acquire a global optimum in prediction performance. Therefore, a greedy algorithm is applied to identify an optimized input combination at each Reynolds number. In each iteration, the pressure signal that provides the largest improvement in prediction performance among the 32 candidates is added to the input. The max iteration number is set to 5 in this study. The optimization results are shown in Fig.\ref{fig:opt_pt}, where two notable features can be observed: (1) the $3^{rd}$-$5^{th}$ iterations are almost collapsed, consistent with the exponential convergence of prediction performance shown in Fig.\ref{fig:f1}b; 
\begin{figure}[h]
     \linespread{1.} 
     \centering
     \includegraphics[width=0.9\linewidth]{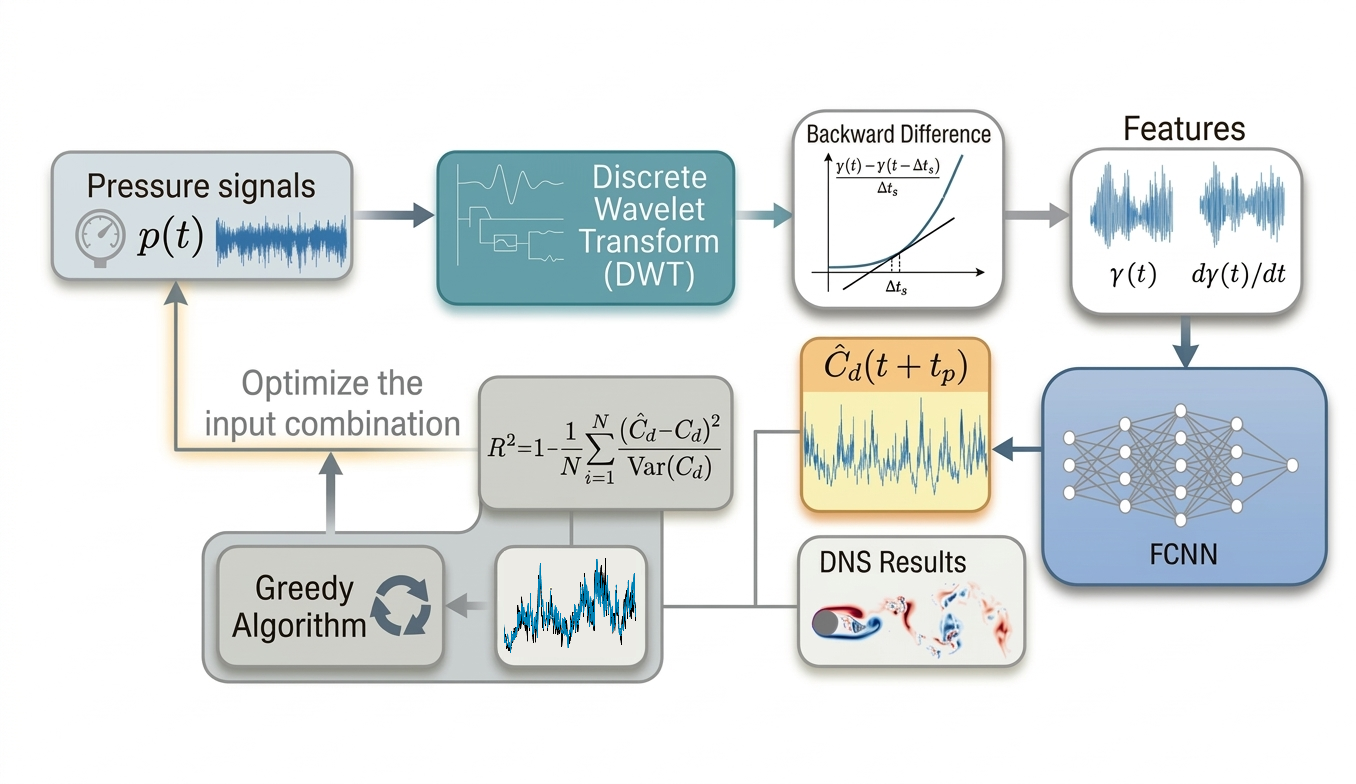}
     \caption{\textbf{Diagram showing the drag prediction in this study.} Data processing for input and optimization iterations are included.}
     \label{fig:prc}
 \end{figure}
(2) the first iteration (shown by the darkest blue symbols) exhibits a pigeon-shaped distribution at all Reynolds numbers, especially at higher Reynolds numbers. This pigeon-shaped distribution reflects the underlying cylinder wake dynamics\cite{williamson_vortex_1996}: the trailing edge interacts with the chaotic wake and is therefore associated with strong temporal gradients in the input signals, while the leading edge experiences a developing laminar state, where the fluctuations are relatively weak and less informative. The optimized combinations of pressure signals at each Reynolds number are summarized in Tab.\ref{tab:opt_pt} and validated by using a shorter training set. The robustness of the optimized sensor locations (also shown in Fig.\ref{fig:f1}a) indicates that these results are physics-driven rather than dependent on the specific setup of the prediction model. Note that the best-performing locations shown in Fig.\ref{fig:f1}a are mirrored for clarity, taking advantage of the cylinder’s symmetry. The procedure of drag prediction is outlined in Fig. \ref{fig:prc}.

\begin{figure}[h]
    \linespread{1.} 
    \centering
    \includegraphics[width=0.8\linewidth]{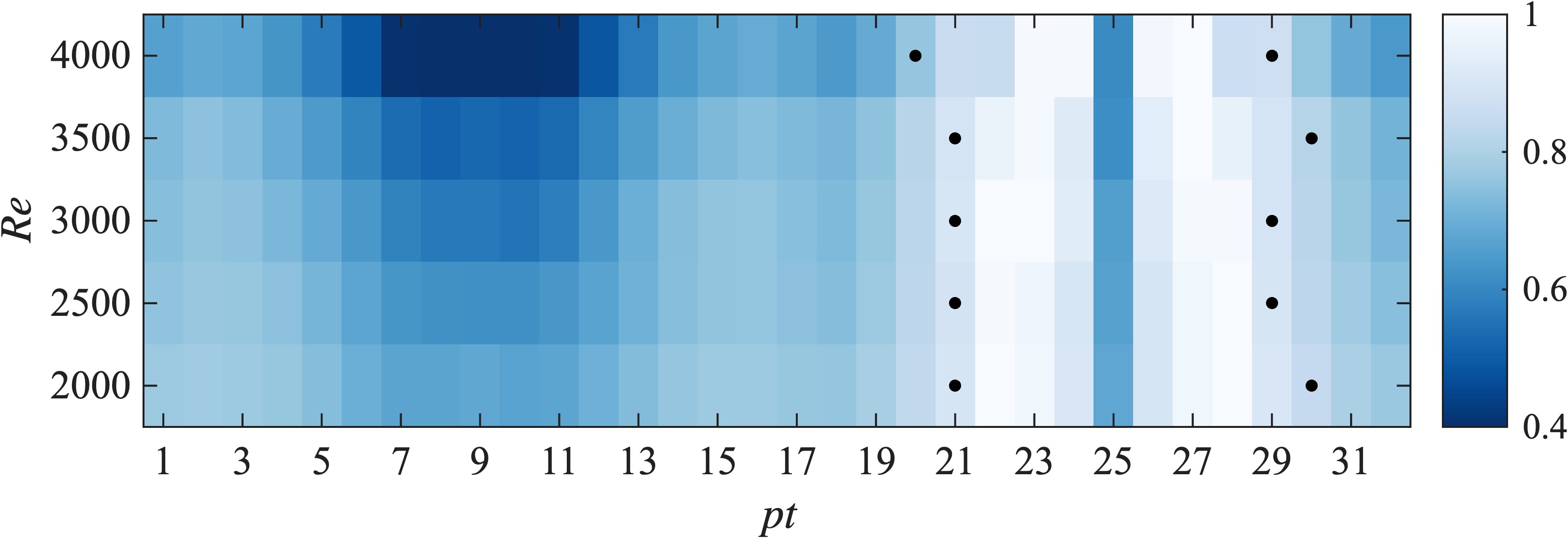}
    \caption{The mutual information between drag coefficient $C_d$ and DWT processed pressure signals ($\gamma$ and $d\gamma/dt)$ at different Reynolds numbers. The colorbar represents the normalized mutual information by its maximum at every Reynolds number. The black symbols show the first two best-performing locations for FCNN's inputs by the greedy algorithm.}
    \label{fig:mi}
\end{figure}

The mutual information \cite{shannon1949mathematical} quantifies the overlap in information content between signals, defined as:
\begin{equation}
    MI(X;Y)=H(X)-H(X|Y),
    \label{eq:mi}
\end{equation}
where $H(X)=-\sum\limits_{x\in X}p(x)\log p(x)$ and $H(X|Y)$  represent the Shannon entropy and the conditional entropy, respectively, for variable sets $X$ and $Y$. The probability density function of $x$ is denoted by $p(x)$. The conditional entropy can be estimated by the joint entropy: $H(X|Y)=H(X,Y)-H(Y)$, where $H(X,Y)=-\sum\limits_{x\in X}\sum\limits_{y\in Y}p(x,y)\log p(x,y)$. For the discrete signals generated by DNS in this study, the Shannon entropy is numerically calculated by the $k$-nearest-neighbour entropy estimator (Kozachenko-Leonenko estimator)\cite{kozachenko1987sample}, with $k=3$ and the $L_\infty$-norm metric (Chebyshev distance). The dataset at each Reynolds number is divided into ten subsets, and the calculation results are averaged. The mutual information $MI([\gamma ~ d\gamma/dt],C_d)$ is normalized by its maximum at every Reynolds number, as shown in Fig.\ref{fig:mi}. The variable set $[\gamma ~ d\gamma/dt]$ corresponds to the FCNN input during the optimization iterations. The optimized results of FCNN's input (black symbols in Fig.\ref{fig:mi}) nearly overlap with the most informative locations, demonstrating that the model captures physically meaningful information in the flow. Consequently, the observed scaling law is not a limitation of the predictive model, but a direct reflection of the intrinsic dynamics in the underlying system.

\begin{figure}[h]
    \linespread{1.} 
    \centering
    \includegraphics[width=1\linewidth]{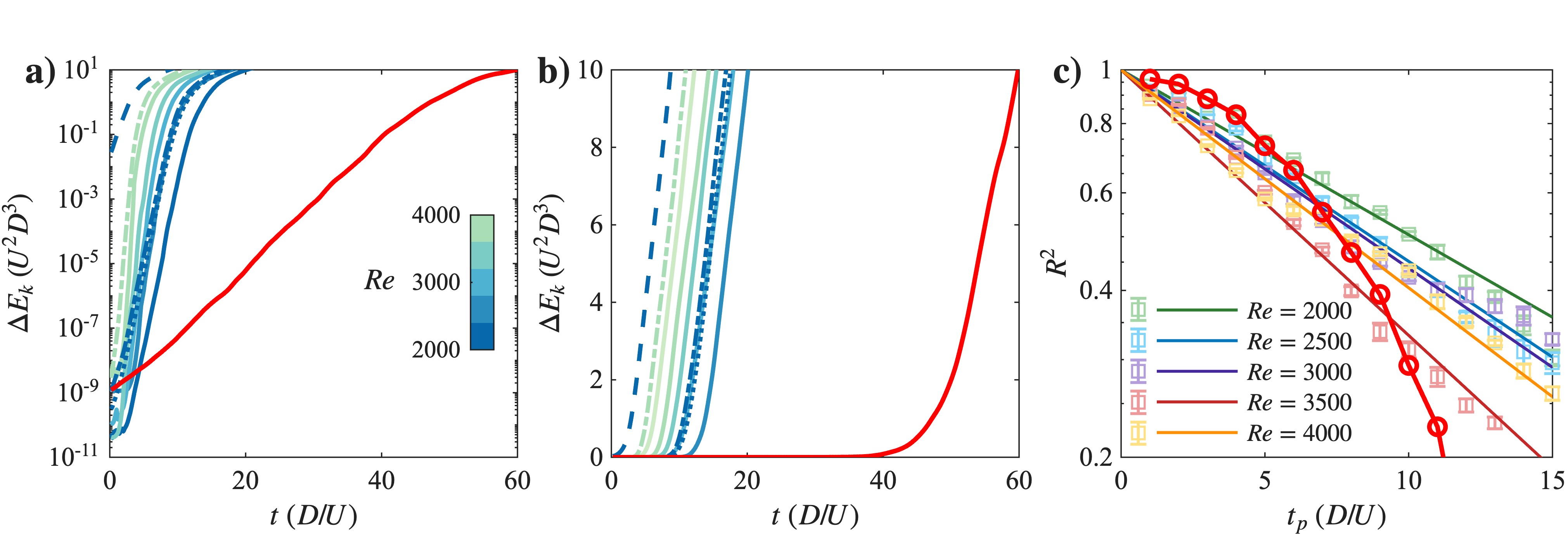}
    \caption{\textbf{A counterexample at $Re=500$.} The data at $Re=500$ are shown as red symbols, while the definitions of the remaining symbols are consistent with those in Fig.\ref{fig:f2} and Fig.\ref{fig:f3}.}
    \label{fig:re500}
\end{figure}

{\fontsize{16}{16}\selectfont \textbf{A Counterexample}}\\
The optimization of pressure signals, the prediction at different forecast horizons, and numerical intervention experiments are implemented again at $Re=500$, and results are shown in Fig. \ref{fig:re500}. Corresponding to the weak chaotic dynamics in \textit{the disordered three dimensionalities regime}\cite{williamson_vortex_1996}, the Lyapunov exponent at $Re=500$ is relatively small yet positive (Fig.\ref{fig:re500}a). The transition from the exponential growth to the linear growth is obviously slower than that indicated by the green-blue series curves in Fig.\ref{fig:re500}b, indicating the absence of an abrupt onset of linear growth. The prediction performance at $Re=500$ initially decreases more slowly than the exponential scaling law would suggest, but followed by a rapid drop (Fig.\ref{fig:re500}c). This difference is reminiscent of the contrast between Weibull and exponential distributions. The former exhibits memory effects, corresponding to the dynamics observed at relatively low Reynolds numbers \cite{mukund_aging_2025}.  

\clearpage
    \bibliography{ref} 
\end{document}